\documentclass[a4paper,11pt]{article}
\usepackage[margin=1.5in]{geometry}
\usepackage{changepage}
\usepackage{cite}
\usepackage[font=small,textfont=it,labelfont=bf]{caption}
\usepackage{amsmath,amssymb,amsfonts}
\usepackage{bm}
\usepackage{subfig}
\usepackage{graphicx}
\usepackage{textcomp}
\usepackage{xcolor}
\usepackage[switch]{lineno}
\usepackage{soul}
\usepackage[strings]{underscore}

\providecommand{\keywords}[1]
{
  \small	
  \textbf{\textit{Keywords---}} #1
}

\title{Real-time ultra-low power {ECG} anomaly detection using an event-driven neuromorphic processor}

\author{
    Felix~Christian~Bauer\\
    \small \textit{aiCTX AG}\\
    \small Zurich, Switzerland\\
    \small felix.bauer@aictx.ai
    \and Dylan~Richard~Muir\\
    \small \textit{aiCTX AG}\\
    \small Zurich, Switzerland\\
    \small dylan.muir@aictx.ai
    \and Giacomo~Indiveri\\
    \small \textit{University of Zurich \& ETH Zurich}\\
    \small Zurich, Switzerland\\
    \small giacomo@ini.uzh.ch
}

\begin{document}

\maketitle

\begin{abstract}
  Accurate detection of pathological conditions in human subjects can be achieved through off-line analysis of recorded biological signals such as electrocardiograms (ECGs).
  However, human diagnosis is time-consuming and expensive, as it requires the time of medical professionals. This is especially inefficient when indicative patterns in the biological signals are infrequent. Moreover, patients with suspected pathologies are often monitored for extended periods, requiring the storage and examination of large amounts of non-pathological data, and entailing a difficult visual search task for diagnosing professionals.
 
  In this work we propose a compact and sub-mW low power neural processing system that can be used to perform on-line and real-time preliminary diagnosis of pathological conditions, to raise warnings for the existence of possible pathological conditions, or to trigger an off-line data recording system for further analysis by a medical professional.
  We apply the system to real-time classification of ECG data for distinguishing between healthy heartbeats and pathological rhythms.
  
 Multi-channel analog ECG traces are encoded as asynchronous streams of binary events and processed using a spiking recurrent neural network operated in a reservoir computing paradigm.
  An event-driven neuron output layer is then trained to recognize one of several pathologies. 
  Finally, the filtered activity of this output layer is used to generate a binary trigger signal indicating the presence or absence of a pathological pattern.
  We validate the approach proposed using a Dynamic Neuromorphic Asynchronous Processor (DYNAP) chip, implemented using a standard 180\,nm CMOS VLSI process, and present experimental results measured from the chip.
\end{abstract}


\section{Introduction}

Cardiovascular diseases are among the main causes of premature deaths worldwide~\cite{naghavi2017}. Some pathological conditions can be diagnosed at an early stage by electrocardiography (ECG), which measures the electrical excitation of the heart. In such cases quick interventions may lead to better outcomes and even save lives~\cite{McMurray_etal12, Gottlieb_etal88, Gottlieb_etal86}.

However, to detect such early indicators in undiagnosed patients it would be necessary to carry out continuous health monitoring over extended periods of time. Until now this has been impractical for several reasons. First of all, long-term monitoring generates large amounts of data, much of which is likely to be non-pathological. Searching such data for indicative segments is a tedious task, which must be performed by trained medical professionals. Furthermore, to reduce inconvenience for patients, recording devices should be portable --- ideally wearable --- and have a long battery life to avoid frequent recharging or  replacement. This is in conflict with the requirement of the device to perform high accuracy recordings and store or transmit the large amounts of recorded data for off-line processing, as these operations typically consume significant energy.

This conflict can be resolved by devices that locally run algorithms to distinguish and filter out irrelevant data continuously and in real-time, without resorting to external- or cloud-computing resources. In this way the amount of data to be stored or transmitted can be drastically reduced and subsequent in-depth analysis by a diagnosing professional is facilitated.

\begin{figure}[htbp]
\centerline{\includegraphics[width=0.75\textwidth]{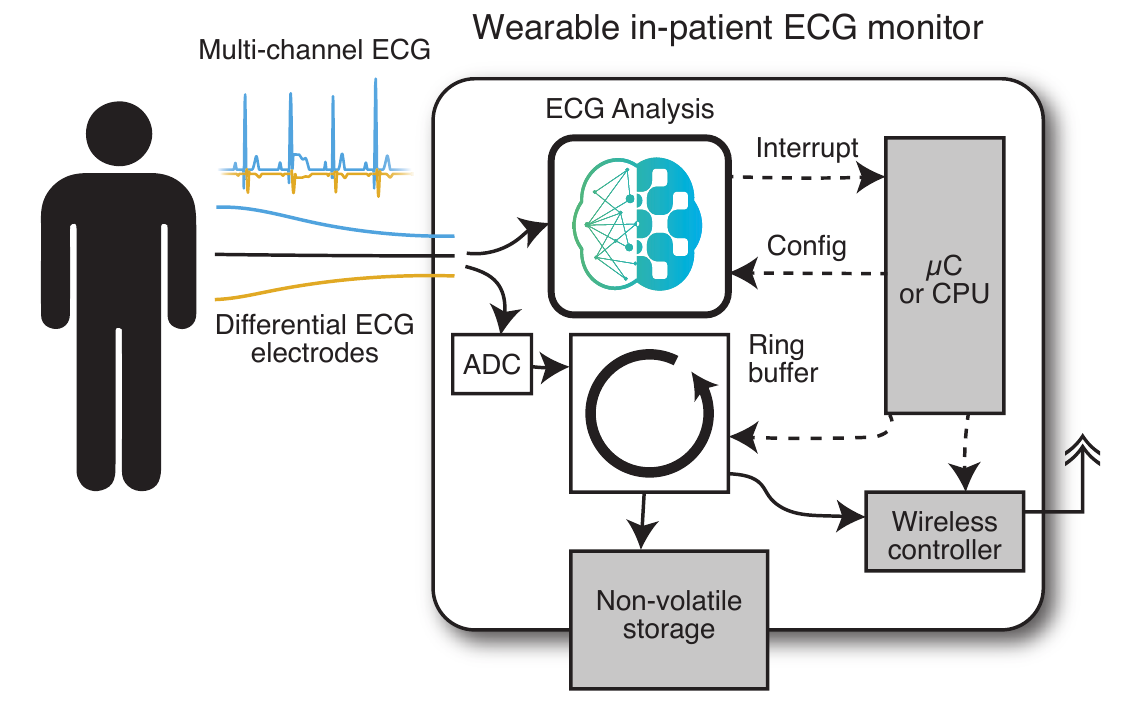}}
\caption{Possible integration of the proposed system into a wearable ECG monitor. A patient's ECG is measured with differential ECG electrodes and continuously analysed. If an indicative pattern is detected, the system wakes up a microcontroller or CPU to record the relevant data onto a non-volatile storage or transmit it to a server via a wireless connection. 
Solid lines: ECG signal pathways; dashed lines: command signals; shaded blocks: components in hibernation mode until a pathological signal is detected.}
\label{fig:ecg_monitor}
\end{figure}

Such a system could, for instance, be part of a wearable ECG monitoring device that is provided to patients with a suspected cardiovascular condition (see Fig.~\ref{fig:ecg_monitor}). The patient's electrocardiogram is continuously measured through sensor electrodes and analyzed by our proposed computing system. In this system, the majority of the system components can be in hibernation for low-power operation (shaded components). If a suspicious ECG rhythm is detected, an interrupt is sent to a microcontroller or CPU, which will cause the ECG signal to be recorded onto non-volatile memory or to be transmitted to a server via a wireless connection. In addition, an alarm may be raised if indicative patterns are detected repeatedly, making timely medical treatment more likely. Such a system is designed to perform only preliminary analysis, but can drastically reduce the diagnostic load on the medical professional.

Automated ECG arrhythmia detection has received much interest in the literature.
Typical approaches involve neural networks~\cite{liu_etal2013, kiranyaz_etal2016, Wang_etal2019}, support vector machines~\cite{ubeyli2007}, fuzzy cognitive maps~\cite{Cardenas_etal2019} or the extraction and analysis of \textit{a priori} defined features~\cite{gradl_etal}. While many of these algorithms achieve detection accuracies of beyond 90 per cent, they are generally not applicable for ultra-low power hardware implementation. Algorithms based on spiking neural networks (SNNs) are suggested to be well-suited for sub milliwatt ECG heart rate~\cite{Das_etal2019} and arrhythmia~\cite{Amirshahi_Hashemi_2019} detection; however these networks are only implemented in simulation on conventional computing hardware.

An alternative cloud computing approach involves sending data to an external, more powerful processor~\cite{marzencki_etal2010, Iliev_etal2019}. In spite of advances in data compression~\cite{mamaghanian_etal2011}, continuous data transmission itself adds considerable power demand to the system. Besides, users may be limited in their mobility if they need to remain within range of a receiving device or of a wireless network.

\begin{figure*}[htbp]
\centerline{\includegraphics[width=0.95\textwidth]{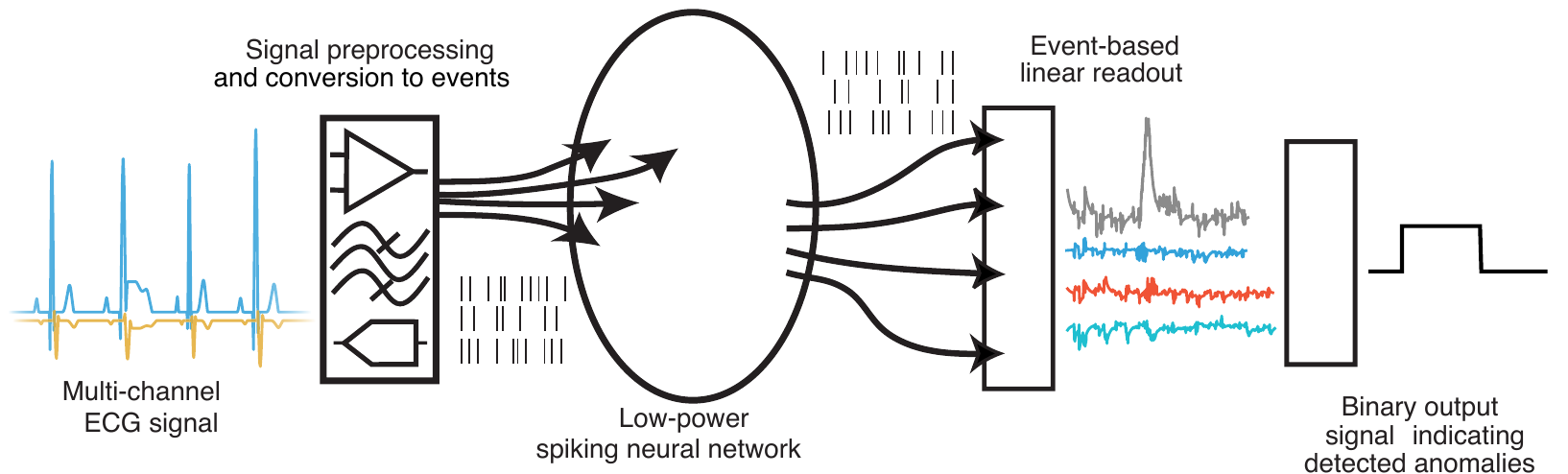}}
\caption{System for real-time ECG anomaly detection. Analog sensor inputs are amplified and filtered in a preprocessing stage and then converted to events using Lebesgue sampling. Resulting event trains are processed by a spiking neural network implemented on a neuromorphic processor. For each type of anomalous pattern a readout unit low-pass filters a weighted sum of neuron firing activities to estimate the likelihood of an anomaly being present in the ECG input. A final binary output signal indicates whether any readout unit is above its detection threshold.}
\label{fig:system-overview}
\end{figure*}

Here we propose a scalable, always-on system architecture for continuous signal processing at microwatt power levels (see Fig.~\ref{fig:system-overview}), which exploits the ultra-low power characteristics of Very Large Scale Integration (VLSI) neuromorphic processors~\cite{Chicca_etal14}. Input is evaluated continuously and in real-time. No segmentation of the signal or synchronization to heartbeats is required.

In the approach proposed, analog signals are sensed, amplified, filtered, and then converted to trains of events using Lebesgue sampling via a sigma-delta encoder~\cite{Corradi_Indiveri15}. The resulting event trains are expanded by a spiking recurrent neural network (sRNN) of leaky integrate-and-fire (LIF) neurons, implemented on the neuromorphic hardware~\cite{Indiveri_etal11}. Neural firing activities are combined and filtered by an event-based linear readout layer that has been trained to detect the presence of specific anomalous patterns in the ECG input. Finally a binary output is generated that indicates the presence or absence of a pathological event.

We validate the architecture proposed by implementing the sRNN on the fully asynchronous, mixed-signal DYNAP-SE device~\cite{Moradi_etal18}, and show experimental results performing real-time ultra-low power analysis of a two-channel ECG signal. In particular, we show that anomalies are detected with high reliability and demonstrate the feasibility of the real-world application of a system as described above. Similar approaches have recently been applied to classify EMG signals with a feed-forward SNN~\cite{Donati_etal2019} and for classification of ECG anomalies with a sRNN~\cite{Corradi_etal2019}, using the same type of hardware.

For other processing stages of the system we refer to existing implementations. Their specifications, together with experimental findings from the DYNAP-SE implementation, allow us to estimate power consumption of a complete system. 

To our best knowledge this work presents the first full-system description of an ECG anomaly detector based on a neuromorphic processor and one of the first implementations of a real-world application on the DYNAP-SE system.

\section{Methods}

\subsection{The DYNAP-SE system}\label{dynapse}
\begin{figure}[htbp]
\centering
\subfloat[]{\includegraphics[width=0.45\textwidth]{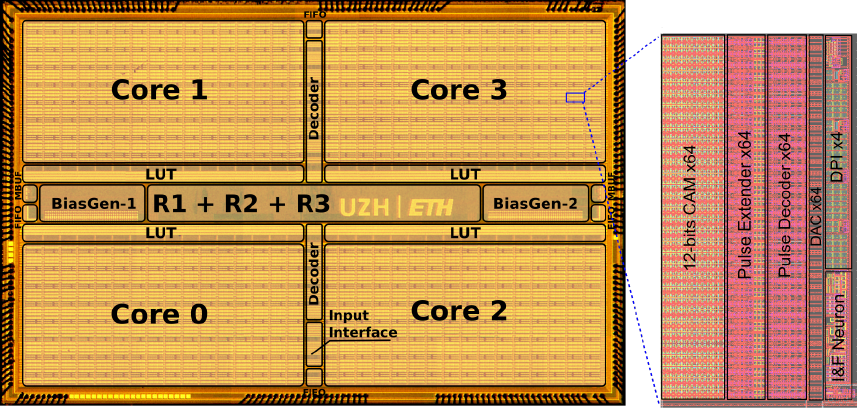}\label{fig:die}}
\hfill
\subfloat[]{\includegraphics[width=0.45\textwidth]{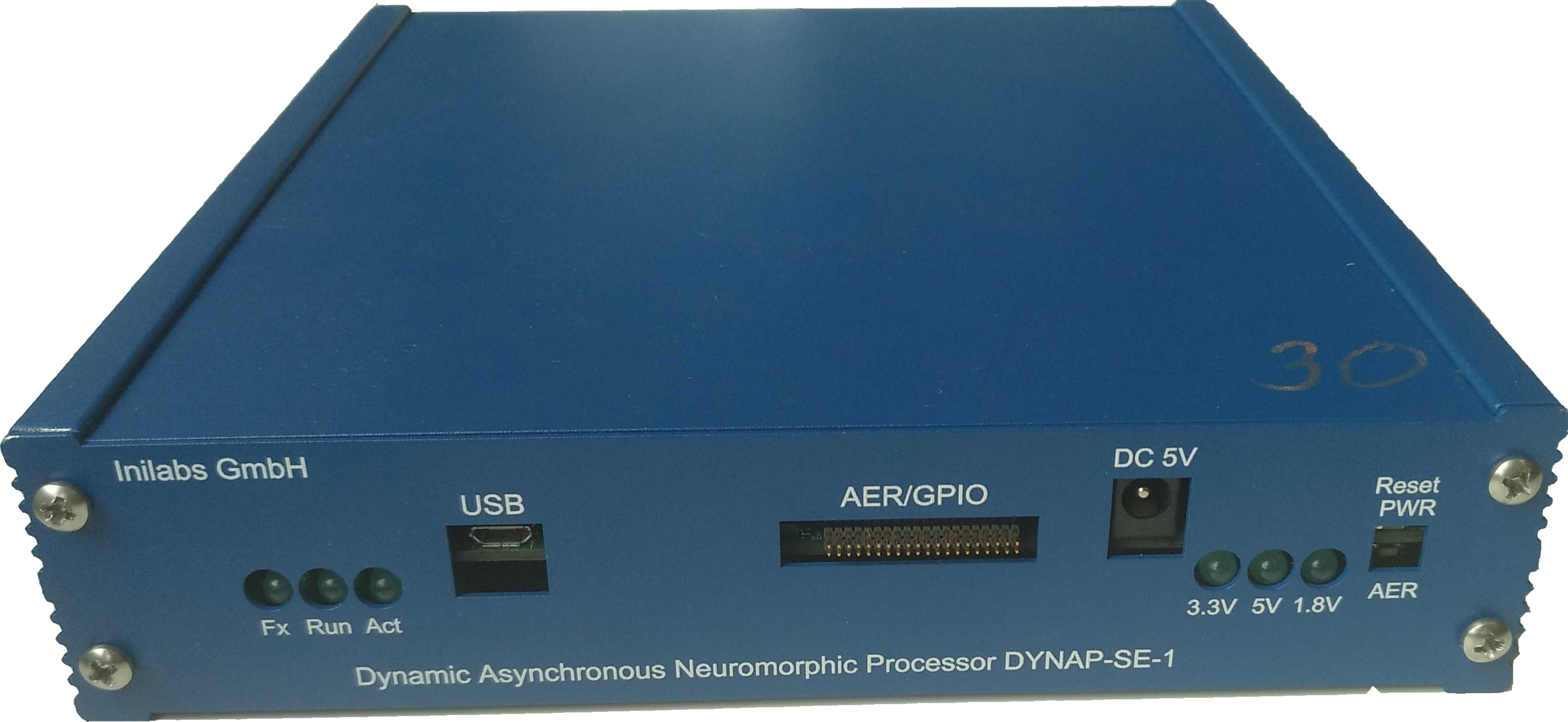}\label{fig:devkit}}
\caption{a) Die photo of the DYNAP-SE multi-core neuromorphic processor~\cite{Moradi_etal18}. The chip is fabricated in standard 180 nm CMOS technology and comprises four cores with 256 adaptive exponential integrate-and-fire neurons each. b) DYNAP-SE development kit with USB interface to a standard PC.}
\label{fig:dynapse}
\end{figure}

The sRNN was implemented on a fully asynchronous, mixed-signal DYNAP-SE chip, first presented and fully characterized in~\cite{Moradi_etal18} (see Fig.~\ref{fig:die}). The DYNAP-SE chip comprises four cores of 256 Adaptive Exponential Integrate-and-Fire (AdExp) neurons, excitatory and inhibitory dynamic synapses implemented using a ``Differential-Pair Integrator'' (DPI) log-domain circuit~\cite{Bartolozzi_Indiveri07a}, and a hierarchical routing architecture for transmitting neural spiking events within cores, among multiple cores and across multiple chips.

Every DYNAP-SE AdExp silicon neuron can subscribe to events from up to 64 presynaptic neurons. For each inter-neuron connection one of four DPI synapse types can be selected, two of which are excitatory and two inhibitory; each with individually tunable characteristics. Neural and synaptic dynamics can be adjusted by setting 25 different circuit parameters, programmed via a temperature-compensated on-chip bias generator~\cite{Delbruck_etal10}, shared by neurons on the same core. This also holds for presynaptic weights for same synapse types. To achieve stronger synaptic weights, multiple connections can be implemented between a pair of neurons, thereby effectively multiplying the weight by the number of connections.

Due to the analog nature of the silicon neurons and their device mismatch, temporal and functional characteristics, such as firing thresholds, synaptic efficacy, and time constants, vary between individual neurons and synapses. This intrinsic in-homogeneity is exploited to introduce variability in neuron parameters and enhance signal dimensionality, as explained in Section~\ref{ss:architecture}.

For this work, a DYNAP-SE development kit was used, which comprises four DYNAP-SE chips and a Field Programmable Gate Array (FPGA), providing a USB interface to a standard PC for configuring circuit parameters, setting up synaptic connections, sending input events and reading out neural firing activity (see in Fig.~\ref{fig:die}).

\subsection{ECG data set and conversion to spikes}\label{dataset}
The set of ECG signals used to evaluate performance in this work is taken from the MIT-BIH Arrhythmia Database~\cite{MIT-BIH}, provided by Massachusetts Institute of Technology and Beth Israel Hospital through PhysioNet~\cite{PhysioNet}. The data set consists of 48 half-hour excerpts from a set of 4000 ambulatory two-channel ECG recordings from 47 subjects. Every ECG rhythm is labeled either as normal or as exhibiting one of 18 anomalous conditions. 23 of the recordings were picked at random while the remaining 25 where selected to include less common anomaly types and parts of low signal quality to challenge arrhythmia detectors. The signal is band-pass filtered between 0.1~Hz and 100~Hz and digitized with a sampling rate of 360~Hz.

\begin{figure}[htbp]
\centerline{\includegraphics[width=0.95\textwidth]{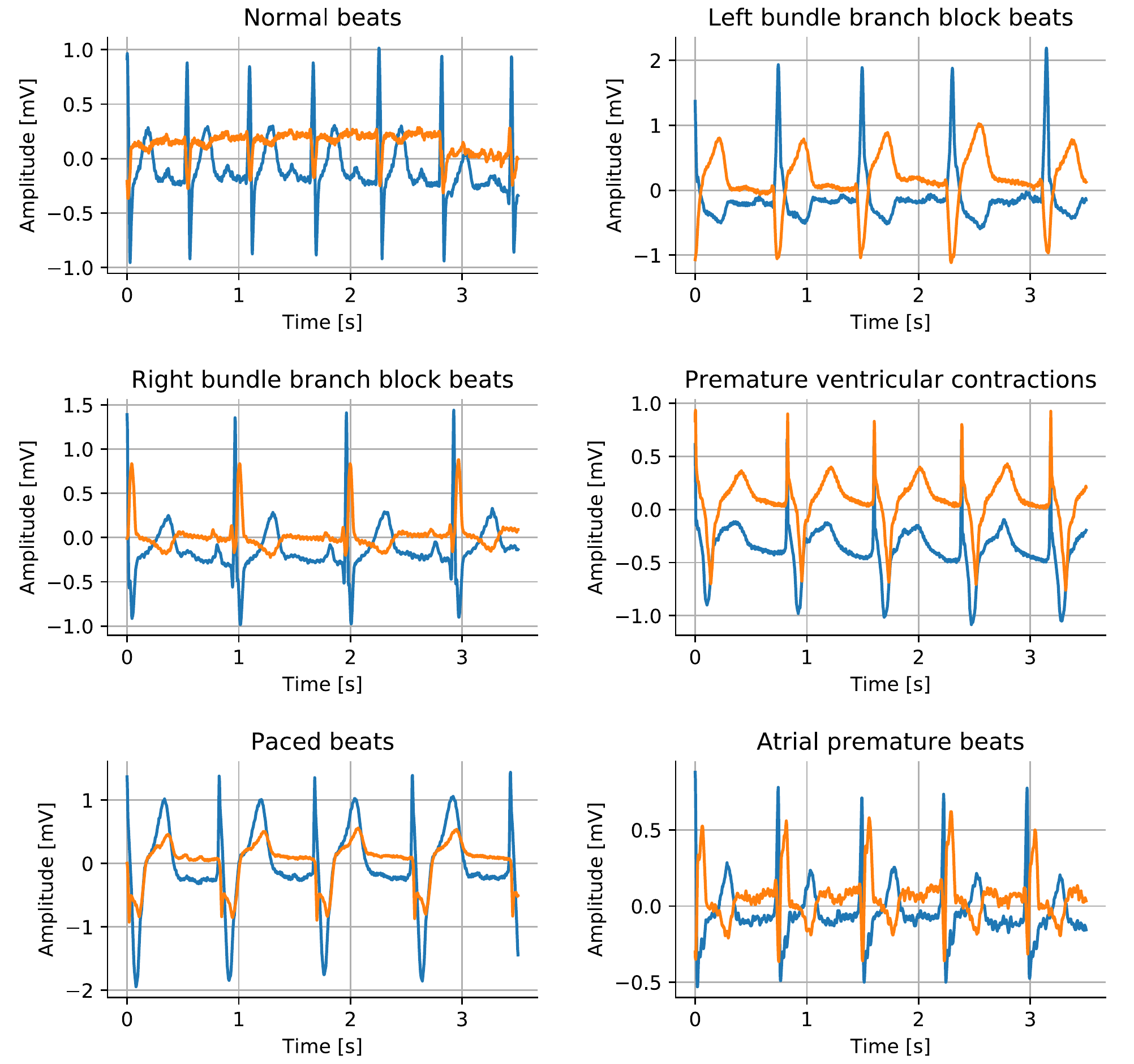}}
\caption{Examples of the ECG signal for normal heartbeats (upper left plot) and each of the five anomaly types used in this work. Blue curves correspond to the first, orange curves to the second channel of the respective recordings.}
\label{fig:beats}
\end{figure}

Some anomalies occur very rarely in the data set. In this work only normal rhythms and the five most prevalent anomaly types were considered, which accounts for 95.6~\% of total recording time. Examples for normal heartbeats and the anomalies used in this work are shown in Fig.~\ref{fig:beats}.

Three disjoint subsets of the input data were used for training, validation and testing of the system. For the training set, recordings were segmented into individual heartbeats. From these beats 30000 were then drawn at random and concatenated to form the training input. For validation and test sets, recordings were split into longer segments, each comprising five to ten contiguous heartbeats of the same label. In a similar fashion as with the training set, segments were then selected at random and concatenated to form the input signals. For all three sets, samples were drawn such that normal rhythms made up about 75~\% of each set and any of the included anomaly types 5~\%. Table~\ref{tab:num_beats} holds the number of beats in the three subsets for each label, Table~\ref{tab:num_segs} the number of continuous anomalous segments in the validation and test sets.

\bgroup
\begin{table}[htbp]
\small
\begin{center}
\begin{tabular}{ |c|c|c|c| } 
 \hline
 \textbf{Beat type}&\multicolumn{3}{c|}{\textbf{Subset}} \\
 \cline{2-4}
 \textbf{(expert label)} & \textbf{Train.} & \textbf{Valid.} & \textbf{Test.} \\
 \hline
 \textit{Normal rhythm} & 22,500 & 2,315 & 1,569 \\ 
 \hline
 \textit{Left bundle branch block beat} & 1,500 & 175 & 104\\ 
 \hline
 \textit{Right bundle branch block beat} & 1,500 & 176 & 103\\ 
 \hline
 \textit{Premature ventricular contraction} & 1,500 & 154 & 100\\ 
 \hline
 \textit{Paced beat} & 1,500 & 153 & 98\\ 
 \hline
 \textit{Atrial premature beat} & 1,500 & 168 & 104\\ 
 \hline
 \textbf{Total} & 30,00 & 3,116 & 2,078\\ 
 \hline
\end{tabular}
\caption{Number of heartbeats per label}
\label{tab:num_beats}
\end{center}
\end{table}

\bgroup
\begin{table}[htbp]
\small
\begin{center}
\begin{tabular}{ |c|c|c| } 
 \hline
 \textbf{Anomaly type}&\multicolumn{2}{c|}{\textbf{Subset}} \\
 \cline{2-3}
 \textbf{(expert label)} & \textbf{Validation} & \textbf{Testing} \\
 \hline
 \textit{Left bundle branch block beat} & 24 & 15\\ 
 \hline
 \textit{Right bundle branch block beat} & 22 & 17\\ 
 \hline
 \textit{Premature ventricular contraction} & 22 & 14\\ 
 \hline
 \textit{Paced beat} & 21 & 16\\ 
 \hline
 \textit{Atrial premature beat} & 22 & 14\\ 
 \hline
 \textbf{Total} & 111 & 76\\ 
 \hline
\end{tabular}
\caption{Number of segments per anomaly type}
\label{tab:num_segs}
\end{center}
\end{table}

ECG signals are converted to trains of events a through sigma-delta encoding scheme~\cite{Corradi_Indiveri15}. For every ECG channel there are two pulse outputs, emitting events either when the input signal increases by a specified amount (up-events) or decreases by a given amount (down-events).

\subsection{sRNN architecture}\label{ss:architecture}
The architecture of the sRNN is illustrated in Fig.~\ref{fig:network} and has been inspired by the paradigm of reservoir computing~\cite{Maass_etal02, Jaeger02}. In these architectures the weights of the recurrent network are initialized randomly and learning takes place only in the output layer, which reads out the state of a hidden layer sampled from the recurrently connected neurons. The network consists of three neuron populations that are implemented on the DYNAP processor: a feed-forward input expansion layer as well as a recurrent excitatory and a non-recurrent inhibitory group that together form the reservoir layer. Random connections and hardware mismatch (see Section~\ref{dynapse}) serve to project the original signal to a high-dimensional state space. Recurrent connections allow for the state to also encode information about the recent past of the input. 

\begin{figure}
\centerline{\includegraphics[width=0.9\textwidth]{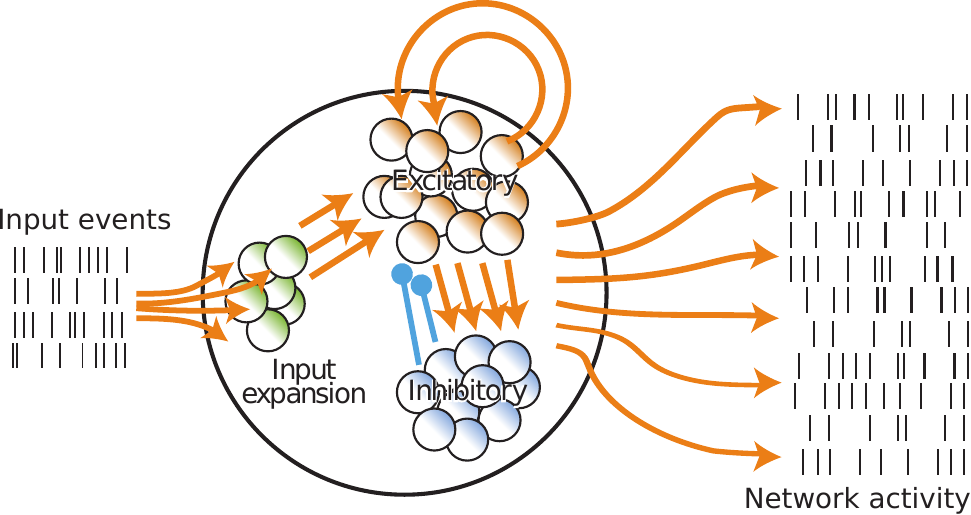}}
\caption{Recurrent sRNN architecture. Dimensionality of the spiking input is expanded by a feed-forward input expansion layer with post-synaptic connections to the excitatory population of a partitioned reservoir layer. The latter has recurrent connections to itself and additionally stimulates a feed-forward inhibitory population that in turn controls activation of the excitatory population.}
\label{fig:network}
\end{figure}

The input expansion layer consists of 128 neurons. Each has a fixed number of 1 to 64 presynaptic excitatory connections, drawn uniformly at random,  to one of the input channels. Together with the mismatch between the hardware neurons, this connection scheme ensures that each neuron responds differently to the input spike trains, therefore increasing the dimensionality of the signal.

The reservoir layer is partitioned into 512 excitatory and 128 inhibitory neurons. Each excitatory neuron receives presynaptic input through 16 excitatory connections from the input expansion layer, 32 recurrent connections from other excitatory neurons of the same layer and 16 connections from the inhibitory population. Each inhibitory neuron holds presynaptic connections to 64 of the excitatory neurons. 

This results in the following connectivities between the three different neuron populations:
\begin{align*}
    & Input \ expansion\text{ to }excitatory: & 12.5 \% \\
    & Excitatory\text{ to }excitatory: & 6.25 \% \\
    & Excitatory\text{ to }inhibitory: & 12.5 \% \\
    & Inhibitory\text{ to }excitatory: & 3.1 \%
\end{align*}
All connections are drawn at random. Multiple connections between the same pair of neurons may exist.

The three above-mentioned neuron populations are implemented on different cores of the neuromorphic processor, which allows setting hardware parameters individually for each and facilitates tuning full-network dynamics. Ideally the network is close to the edge of chaos~\cite{Langton90}, where dynamics are rich in response to an input signal but return to a stable state after external stimulation ends.

\subsection{Learning and readout}
Spiking activities of all hardware neurons are constantly monitored with a PC. Spike trains are low-pass filtered by convolution with an exponential kernel \(\kappa(t) = \exp{-\frac{t}{\tau_{out}}}\) with time constant \(\tau_{out} = 0.175 \ s\):
\[ \vec{x}(t) = \big(\vec{s} * \kappa \big) (t) \]
Here, \(\vec{s}(t)\) is a vector containing the spike trains of all hardware neurons and \(\vec{x}(t)\) a vector of the filtered spike trains. 

For each anomaly type in the input signal one readout unit is trained to detect it by taking a weighted sum of the filtered spiking activities. If this sum exceeds a fixed threshold, the anomaly counts as detected. Training of readout weights and thresholds is done independently for each anomaly type.

Readout weights are trained using linear least squares approximation, such that for all readout units \(i\)
\[ \|X \cdot \vec{w}^{(i)} - \vec{y}^{(i)} \| \]
is minimized. Here, \(\vec{w}^{(i)}\) is a vector holding input weights for unit \(i\). The rows of matrix \(X\) are the filtered spike trains \(\vec{x}\), sampled over time. The vector \(\vec{y}^{(i)}\) is the target at corresponding time points. It is 1 whenever an anomaly of type \(i\) is present in the input signal and 0 otherwise.

During a validation run, a subset of the ECG data is used to find optimal detection thresholds for the readout units. Thresholds are chosen such that the number of false negatives for the respective anomaly and the number of false positives are minimised over the validation set.

\section{Results}
\subsection{Network performance}
As described in Section~\ref{dataset} the input signal to the network is assembled from segments of contiguous heartbeats. Within a segment, beats share the same label, which is either \textit{normal} or one of the five considered anomalous patterns. The task for the sRNN is to indicate the presence of anomalous conditions in the ECG. Classification of the specific anomaly type is not required. Network output can be interpreted as binary, being positive whenever any of the readout units indicates a detected anomaly and negative otherwise.

Network performance is quantified by means of \(sensitivity\), \(specificity\), \(PPV\) (positive predictive value), and \(NPV\) (negative predictive value), which are defined as follows:
\begin{align*}
    sensitivity &= \frac{\# \ \text{true positives}}{\# \ \text{true positives} + \# \ \text{false negatives}} \\
    specificity &= \frac{\# \ \text{true negatives}}{\# \ \text{true negatives} + \# \ \text{false positives}} \\
    PPV &= \frac{\# \ \text{true positives}}{\# \ \text{true positives} + \# \ \text{false positives}} \\
    NPV &= \frac{\# \ \text{true negatives}}{\# \ \text{true negatives} + \# \ \text{false negatives}} \\
\end{align*}

A true positive (false negative) is any segment of anomalous heartbeats during which the presence of an anomaly is (is not) indicated by any of the readout units. A true negative (false positive) is any normal heartbeat during which the network does not (does) indicate the presence of an anomaly. Because only the combined output of all readout units is considered, a unit indicating the presence of an anomaly which does not correspond to its own target label still counts as correct detection.

Overall, 91.3~\% of anomalous segments in the test set were detected. Of the normal heartbeats, 2.4~\% were falsely indicated as being anomalous. This means that for a normal ECG input, on average there would be a false positive every 31.5 seconds.

As shown in Table~\ref{tab:performance}, for three of the five presented anomaly types the network detected all of the anomalous segments reliably, for another type sensitivity was 88~\%. Only for one anomaly, \textit{premature ventricular contraction}, detection was significantly less reliable with 71.43~\%. There is no apparent correlation between individual subjects or ECG recordings and sensitivity for this anomaly. 

Specificity is above 99~\% for all anomaly types but particularly high for \textit{left bundle branch block beat} and \textit{atrial premature beat}, which resulted in only two and one false positives over the whole test set, respectively. 

Values for PPV are lower than the corresponding sensitivities, the lowest being 45.5~\% for \textit{premature ventricular contraction}, whereas NPV is high for all conditions.

\begin{table*}[t]
\begin{adjustwidth}{-0.5in}{}
\small
\begin{center}
\begin{tabular}{ |c|c|c|c|c|c| } 
 \hline
 \textbf{Anomaly type} & \textbf{Sensitivity} & \textbf{Specificity} & \textbf{PPV} & \textbf{NPV} & \textbf{MTBFP$^{\mathrm{a}}$}  \\
 \textbf{(expert label)} & \textbf{(in \%)} & \textbf{(in \%)} & \textbf{(in \%)} & \textbf{(in \%)} & \textbf{(in seconds)} \\
 \hline
 \textit{Left bundle branch block beat} & 100.0 & 99.87 & 88.24 & 100.0 & 582.8 \\ 
 \hline
 \textit{Right bundle branch block beat} & 88.24 & 99.36 & 60.0 & 99.87 & 116.56\\ 
 \hline
 \textit{Premature ventricular contraction} & 71.43 & 99.24 & 45.45 & 99.74 & 97.1\\ 
 \hline
 \textit{Paced beat} & 100.0 & 99.24 & 57.14 & 100.0 & 97.1\\ 
 \hline
 \textit{Atrial premature beat} & 100.0 & 99.94 & 93.33 & 100.0 & 1165.6\\ 
 \hline
 \textbf{Overall} & 92.11 & 97.6 & 65.42 & 99.61 & 31.5\\
\hline
\multicolumn{6}{l}{\footnotesize $^{\mathrm{a}}$Mean time between false positives during normal ECG input}
\end{tabular}
\caption{Anomaly detection performance on test set}
\label{tab:performance}
\end{center}
\end{adjustwidth}
\end{table*}

\subsection{Power consumption}

Based on the figures provided in~\cite{Moradi_etal18}, the firing activity and architecture of the sRNN on the DYNAP-SE chip translate into a dynamic power consumption of 286.1~\(\mu\)W. Assuming a static power draw of around 230~\(\mu\)W, total power consumption amounts to 516.1~\(\mu\)W. Together with the power estimates for those parts of the system that were simulated on a desktop computer (see Section~\ref{ssec:feasibility}), the total power consumption of the proposed system is less than 722.1~\(\mu\)W. An overview of the power figures for different components can be found in Table~\ref{tab:power}.

\bgroup
\begin{table*}[t]
\small
\begin{center}
\begin{tabular}{ p{5mm}|c|c|l } 
 \cline{2-3}
 & \textbf{Component} & \textbf{Power} \\
 \cline{2-3}
 & \textit{DYNAP-SE total} & 516.1 \(\mu\)W \\ 
 \cline{2-3}
 & \textit{Amplifiers} & \( < 2 \ \times \ 50 \ \mu\)W$^{\mathrm{a}}$ \\ 
 \cline{2-3}
 & \textit{Filters} & \( < 4 \ \times\ 4 \ \mu\)W$^{\mathrm{b}}$ \\ 
 \cline{2-3}
 & \textit{Readout} & \( < 90 \ \mu\)W \\ 
 \cline{2-3}
 & \textbf{Total} & \(\bm{< 722.1 \ \mu}\)\textbf{W} \\
\cline{2-3}
\multicolumn{4}{l}{\footnotesize $^{\mathrm{a}}$One amplifier per ECG channel} \\
\multicolumn{4}{l}{\footnotesize $^{\mathrm{b}}$One low-pass and one notch filter per ECG channel}
\end{tabular}
\caption{Power consumption}
\label{tab:power}
\end{center}
\end{table*}

\section{Discussion}

\subsection{Performance metric}
In this work network performance is evaluated by means of two key values: sensitivity and specificity, which quantify the ability to detect anomalous patterns in the ECG input on the one hand, and on the other hand proneness to produce erroneous detection events.

In the data set, anomalous patterns usually do not occur in an isolated fashion but appear multiple times in close succession. Therefore missing one pathological heartbeat is not problematic as long as others are detected correctly. Hence, sensitivity analysis is done over segments of rhythms with the same label and any detection counts for the whole segment. However, even for the same anomaly type there may be variations between individual beats, in particular for recordings from different subjects. The network might be able to detect the anomaly with high reliability for some subjects and for others not. To make sure that such a scenario does not go undetected, segments always consist of contiguous heartbeats from single recordings.

The ratio of erroneous detection events and number of normal heartbeats provides an easy way to quantify specificity. Together with the combined duration of all normal beats it can be easily translated into the expected mean time between two false positives for an input that is free of anomalies, a number that gives an intuitive idea of specificity.

\subsection{Network performance}
The network reliably detects most of the presented anomalies, except for \textit{premature ventricular contraction}. There is no apparent correlation between individual subjects and sensitivity for this label and it seems that this pattern is particularly difficult for the network to detect, indicating that performance depends to some extent on the type of anomaly that is to be found. 

\iffalse --TODO: integrate into following paragraph-- Relatively low PPVs and high NPVs suggest that on the one hand not every detection necessarily corresponds to a true anomaly, ...professional should validate positive results. On the other hand, negative results can be trusted with higher confidence. Nevertheless it should be noted that due to the way true positives and false negatives are defined here, their numbers are lower than those for true negatives and false positives, which results in a disproportionally small PPV and large NPV.\fi

Which level of detection accuracy is adequate strongly depends on the individual application. For expert systems that autonomously classify heartbeats, high precision is crucial. On the other hand, in an assisted diagnosis scenario where the system only serves as a filter prior to diagnosis by a human expert, precision requirements may be reduced. For instance, long-term monitoring bio-signals from out-patients traditionally generate large amounts of data which raises challenges that can be alleviated by the suggested system: Diagnosing professionals can be directed to relevant sections without having to sift through the complete recording. The amount of data that needs to be stored and further processed can be reduced by filtering out irrelevant non-pathological sequences. This also allows for more compact devices with long battery life.

Regarding mean time between false positives (MTBFP) and specificity, overall values are lower than for individual anomalies. The reason is that taking into account detection events for all classes simultaneously implies that false positives will accumulate over classes. PPV is slightly low for some anomalies, suggesting that positive results are not to be interpreted as definite diagnosis but rather serve to initiate and support further analysis. It should be noted however, that because true positives and false negatives are defined with respect to segments of multiple heartbeats, their numbers are lower than those for true negatives and false positives, which refer to individual heartbeats. This results in a disproportionally small PPV.

The chosen trade off between accuracy and MTBFP rate can depend on the precise use case. For example, a diagnosing professional can analyze an ECG much more efficiently if only two instead of 70 to 80 rhythms per recorded minute need to be considered. In other scenarios, sensitivity may be weighed up against the number of false alarms by changing detection thresholds. Furthermore, specificity can be improved by only searching for specific anomaly types, e.g. if a specific pathological state is suspected in advance. For example, if only \textit{atrial premature beats} are considered, MTBFP is above 19 minutes with the current setup.

Nevertheless, promising techniques have been proposed that may improve performance significantly, but whose implementation would have exceeded the scope of this work: Instead of using random recurrent connections, network dynamics can be tailored to a given task by spectral analysis of the network~\cite{aceituno_17} or techniques for balancing excitation and inhibition~\cite{deneve_13, Alemi_etal2018}, in-reservoir learning algorithms such as~\cite{wilten_clopath17}, or by using more traditional approaches like Backpropagation Through Time (BPTT)~\cite{Rumelhart_etal86a} to train a non-spiking recurrent neural network and using a transfer algorithm as in~\cite{he2019}.

Finally, while the linear regression algorithm used in this work allows for efficient training of the readout weights, other learning models such as support-vector machines (SVMs)~\cite{Cortes_Vapnik95} may yield more suitable weights for this task. It may even be considered replacing the layer of readout units by a multilayer perceptron (MLP) trained with backpropagation~\cite{rumelhart1985}.

\subsection{Feasibility of preprocessing and readout stages}\label{ssec:feasibility}
While the spiking neural network in this work was implemented on neurmorphic hardware, other parts of the system, namely the conversion from analog input to events, the readout stage and the generation of  the binary output signal were simulated on a desktop computer. Furthermore, the ECG signal that was used had already been amplified and filtered, so that no preprocessing was necessary. In the following we will argue that an implementation of these stages and therefore of the full proposed system is feasible. 

\subsubsection{Preprocessing and conversion to events}
Typical raw ECG sensor signals have an amplitude of a few millivolts and are generally amplified and noise-filtered. Suitable low-power, low-noise amplifiers have been proposed, for instance, in~\cite{yang_etal11} and~\cite{ghamati_nejad13}, each with a power consumption of less than 50~\(\mu\)W. Similarly, lowpass and notch filters are proposed in~\cite{kumar2016} and~\cite{yehoshuva_etal16}, requiring less than 4~\(\mu\)W power. Implementations of asynchronous sigma-delta encoders that efficiently convert the analog signal to events are described in~\cite{Corradi_Indiveri15} and~\cite{Lichtsteiner_etal08}.

\subsubsection{Readout}
The readout units have not been set up as neurons on the neuromorphic processor for two reasons: the limited fan-in of the hardware neurons and the fact that all presynaptic weights of a neuron are the same for a given synapse type.

The fan-in is critical insofar as a readout unit ideally has access to the firing activity of all neurons in the hidden layers of the network. However, it is certainly possible to design neuromorphic processors with readout units that can subscribe to significantly larger numbers of presynaptic neurons. One example is the Reconfigurable On-line Learning Spiking Neuromorphic Processor (ROLLS)~\cite{Qiao_etal15}, which is similar to the DYNAP-SE used in this work and can be configured such that a single neuron has up to 130k synapses. Furthermore, on an algorithmical level the required fan-in can be reduced by variable selection methods and regularization rules that encourage a large number of zero-weights, such as LASSO~\cite{tibshirani96}.

Regarding weight quantization, synapses with individually tunable strengths and a precision of a few bits have been implemented in neuromorphic devices like the Dynap-SEL~\cite{Thakur_etal18}. Although it may not suffice to quantize previously trained high-precision weights, there are learning algorithms such as~\cite{zhu2016},~\cite{hubara2017} and~\cite{Helwegen_etal19} that are designed to find suitable low-precision weights. Another potential solution is the application of transfer methods as suggested in~\cite{he2019}. 

Assuming similar power figures as for DYNAP-SE, we estimate that an implementation of a readout as described above, that produces a binary output indicating whether one of the detection thresholds is exceeded, would consume less than 90~\(\mu\)W.

\section{Conclusion}
In this work we propose an always-on ultra-low power system that uses asynchronous neuromorphic hardware to perform real-time anomaly detection on a multi-channel ECG signal with a mean power consumption below one milliwatt. We implement a spiking recurrent neural network on a DYNAP-SE chip and demonstrate that it is able to reliably detect anomalous patterns in the ECG. Remaining parts of the proposed system can be realized with existing circuitry as part of a wearable health monitoring device.

\section*{Acknowledgment}
This work is supported in part by EU-H2020 grant NeuRAM3 Cube (NEUral computing aRchitectures in Advanced Monolithic 3D-VLSI nano-technologies); by H2020 FET Proactive grant SYNCH (824162); by H2020 ECSEL grant TEMPO (826655); by the European Research Council under the Grant Agreement No. 724295 (NeuroAgents); and by aiCTX AG. Mr Bauer and Dr Muir performed this work as part of their duties at aiCTX AG.

\bibliographystyle{IEEEtran}
\bibliography{library}

\begin{thebibliography}{10}
\providecommand{\url}[1]{#1}
\csname url@samestyle\endcsname
\providecommand{\newblock}{\relax}
\providecommand{\bibinfo}[2]{#2}
\providecommand{\BIBentrySTDinterwordspacing}{\spaceskip=0pt\relax}
\providecommand{\BIBentryALTinterwordstretchfactor}{4}
\providecommand{\BIBentryALTinterwordspacing}{\spaceskip=\fontdimen2\font plus
\BIBentryALTinterwordstretchfactor\fontdimen3\font minus
  \fontdimen4\font\relax}
\providecommand{\BIBforeignlanguage}[2]{{%
\expandafter\ifx\csname l@#1\endcsname\relax
\typeout{** WARNING: IEEEtran.bst: No hyphenation pattern has been}%
\typeout{** loaded for the language `#1'. Using the pattern for}%
\typeout{** the default language instead.}%
\else
\language=\csname l@#1\endcsname
\fi
#2}}
\providecommand{\BIBdecl}{\relax}
\BIBdecl

\bibitem{naghavi2017}
M.~Naghavi, A.~A. Abajobir, C.~Abbafati, K.~M. Abbas, F.~Abd-Allah, S.~F.
  Abera, V.~Aboyans, O.~Adetokunboh, A.~Afshin, A.~Agrawal \emph{et~al.},
  ``Global, regional, and national age-sex specific mortality for 264 causes of
  death, 1980--2016: a systematic analysis for the global burden of disease
  study 2016,'' \emph{The Lancet}, vol. 390, no. 10100, pp. 1151--1210, 2017.

\bibitem{McMurray_etal12}
A.~F. Members, J.~J. McMurray, S.~Adamopoulos, S.~D. Anker, A.~Auricchio,
  M.~B{\"o}hm, K.~Dickstein, V.~Falk, G.~Filippatos, C.~Fonseca \emph{et~al.},
  ``{ESC} guidelines for the diagnosis and treatment of acute and chronic heart
  failure 2012: The task force for the diagnosis and treatment of acute and
  chronic heart failure 2012 of the european society of cardiology. developed
  in collaboration with the heart failure association ({HFA}) of the {ESC},''
  \emph{European journal of heart failure}, vol.~14, no.~8, pp. 803--869, 2012.

\bibitem{Gottlieb_etal88}
\BIBentryALTinterwordspacing
S.~O. Gottlieb, S.~H. Gottlieb, S.~C. Achuff, R.~Baumgardner, E.~D. Mellits,
  M.~L. Weisfeldt, and G.~Gerstenblith, ``{Silent Ischemia on Holter Monitoring
  Predicts Mortality in High-Risk Postinfarction Patients},'' \emph{JAMA}, vol.
  259, no.~7, pp. 1030--1035, 02 1988. [Online]. Available:
  \url{https://doi.org/10.1001/jama.1988.03720070030029}
\BIBentrySTDinterwordspacing

\bibitem{Gottlieb_etal86}
\BIBentryALTinterwordspacing
S.~O. Gottlieb, M.~L. Weisfeldt, P.~Ouyang, E.~D. Mellits, and G.~Gerstenblith,
  ``Silent ischemia as a marker for early unfavorable outcomes in patients with
  unstable angina,'' \emph{New England Journal of Medicine}, vol. 314, no.~19,
  pp. 1214--1219, 1986, pMID: 2871485. [Online]. Available:
  \url{https://doi.org/10.1056/NEJM198605083141903}
\BIBentrySTDinterwordspacing

\bibitem{liu_etal2013}
\BIBentryALTinterwordspacing
S.-H. Liu, D.-C. Cheng, and C.-M. Lin, ``Arrhythmia identification with
  two-lead electrocardiograms using artificial neural networks and support
  vector machines for a portable {ECG} monitor system,'' \emph{Sensors},
  vol.~13, no.~1, pp. 813--828, 2013. [Online]. Available:
  \url{https://www.mdpi.com/1424-8220/13/1/813}
\BIBentrySTDinterwordspacing

\bibitem{kiranyaz_etal2016}
S.~{Kiranyaz}, T.~{Ince}, and M.~{Gabbouj}, ``Real-time patient-specific {ECG}
  classification by {1-D} convolutional neural networks,'' \emph{IEEE
  Transactions on Biomedical Engineering}, vol.~63, no.~3, pp. 664--675, March
  2016.

\bibitem{Wang_etal2019}
N.~{Wang}, J.~{Zhou}, G.~{Dai}, J.~{Huang}, and Y.~{Xie}, ``Energy-efficient
  intelligent {ECG} monitoring for wearable devices,'' \emph{IEEE Transactions
  on Biomedical Circuits and Systems}, pp. 1--1, 2019.

\bibitem{ubeyli2007}
\BIBentryALTinterwordspacing
E.~D. Übeyli, ``{ECG} beats classification using multiclass support vector
  machines with error correcting output codes,'' \emph{Digital Signal
  Processing}, vol.~17, no.~3, pp. 675 -- 684, 2007. [Online]. Available:
  \url{http://www.sciencedirect.com/science/article/pii/S1051200406001941}
\BIBentrySTDinterwordspacing

\bibitem{Cardenas_etal2019}
O.~A. {Cárdenas}, L.~M. {Flores Nava}, F.~G. {Castañeda}, and J.~A. {Moreno
  Cadenas}, ``{ECG} arrhythmia classification based on fuzzy cognitive maps,''
  in \emph{2019 16th International Conference on Electrical Engineering,
  Computing Science and Automatic Control ({CCE})}, Sep. 2019, pp. 1--4.

\bibitem{gradl_etal}
S.~{Gradl}, P.~{Kugler}, C.~{Lohmüller}, and B.~{Eskofier}, ``Real-time {ECG}
  monitoring and arrhythmia detection using android-based mobile devices,'' in
  \emph{2012 Annual International Conference of the IEEE Engineering in
  Medicine and Biology Society}, Aug 2012, pp. 2452--2455.

\bibitem{Das_etal2019}
\BIBentryALTinterwordspacing
A.~Das, P.~Pradhapan, W.~Groenendaal, P.~Adiraju, R.~T. Rajan, F.~Catthoor,
  S.~Schaafsma, J.~L. Krichmar, N.~Dutt, and C.~V. Hoof, ``Unsupervised
  heart-rate estimation in wearables with liquid states and a probabilistic
  readout,'' \emph{Neural Networks}, vol.~99, pp. 134 -- 147, 2018. [Online].
  Available:
  \url{http://www.sciencedirect.com/science/article/pii/S0893608017303003}
\BIBentrySTDinterwordspacing

\bibitem{Amirshahi_Hashemi_2019}
A.~{Amirshahi} and M.~{Hashemi}, ``{ECG} classification algorithm based on
  {STDP} and {R-STDP} neural networks for real-time monitoring on ultra
  low-power personal wearable devices,'' \emph{IEEE Transactions on Biomedical
  Circuits and Systems}, pp. 1--1, 2019.

\bibitem{marzencki_etal2010}
M.~Marzencki, K.~Tavakolian, Y.~Chuo, B.~Hung, P.~Lin, and B.~Kaminska,
  ``Miniature wearable wireless real-time health and activity monitoring system
  with optimized power consumption,'' \emph{Journal of Medical and Biological
  Engineering}, vol.~30, pp. 227--235, 01 2010.

\bibitem{Iliev_etal2019}
I.~T. {Iliev}, I.~I. {Jekova}, S.~D. {Tabakov}, K.~G. {Koshtikova}, and S.~T.
  {Iovev}, ``Telemetry of hospitalized high-risk patients with cardiovascular
  diseases,'' in \emph{2019 IEEE XXVIII International Scientific Conference
  Electronics (ET)}, Sep. 2019, pp. 1--4.

\bibitem{mamaghanian_etal2011}
H.~{Mamaghanian}, N.~{Khaled}, D.~{Atienza}, and P.~{Vandergheynst},
  ``Compressed sensing for real-time energy-efficient {ECG} compression on
  wireless body sensor nodes,'' \emph{IEEE Transactions on Biomedical
  Engineering}, vol.~58, no.~9, pp. 2456--2466, Sep. 2011.

\bibitem{Chicca_etal14}
E.~Chicca, F.~Stefanini, C.~Bartolozzi, and G.~Indiveri, ``Neuromorphic
  electronic circuits for building autonomous cognitive systems,''
  \emph{Proceedings of the {IEEE}}, vol. 102, no.~9, pp. 1367--1388, 9 2014.

\bibitem{Corradi_Indiveri15}
F.~Corradi and G.~Indiveri, ``A neuromorphic event-based neural recording
  system for smart brain-machine-interfaces,'' \emph{Biomedical Circuits and
  Systems, {IEEE} Transactions on}, vol.~9, no.~5, pp. 699--709, 2015.

\bibitem{Indiveri_etal11}
\BIBentryALTinterwordspacing
G.~Indiveri, B.~Linares-Barranco, T.~Hamilton, A.~van Schaik,
  R.~Etienne-Cummings, T.~Delbruck, S.-C. Liu, P.~Dudek, P.~H{\"a}fliger,
  S.~Renaud, J.~Schemmel, G.~Cauwenberghs, J.~Arthur, K.~Hynna, F.~Folowosele,
  S.~Saighi, T.~Serrano-Gotarredona, J.~Wijekoon, Y.~Wang, and K.~Boahen,
  ``Neuromorphic silicon neuron circuits,'' \emph{Frontiers in Neuroscience},
  vol.~5, pp. 1--23, 2011. [Online]. Available:
  \url{http://www.frontiersin.org/Neuromorphic_Engineering/10.3389/fnins.2011.00073/abstract}
\BIBentrySTDinterwordspacing

\bibitem{Moradi_etal18}
S.~Moradi, N.~Qiao, F.~Stefanini, and G.~Indiveri, ``A scalable multicore
  architecture with heterogeneous memory structures for dynamic neuromorphic
  asynchronous processors ({DYNAPs}),'' \emph{Biomedical Circuits and Systems,
  {IEEE} Transactions on}, vol.~12, no.~1, pp. 106--122, Feb. 2018.

\bibitem{Donati_etal2019}
E.~{Donati}, M.~{Payvand}, N.~{Risi}, R.~{Krause}, and G.~{Indiveri},
  ``Discrimination of {EMG} signals using a neuromorphic implementation of a
  spiking neural network,'' \emph{IEEE Transactions on Biomedical Circuits and
  Systems}, vol.~13, no.~5, pp. 795--803, Oct 2019.

\bibitem{Corradi_etal2019}
F.~{Corradi}, S.~{Pande}, J.~{Stuijt}, N.~{Qiao}, S.~{Schaafsma},
  G.~{Indiveri}, and F.~{Catthoor}, ``{ECG}-based heartbeat classification in
  neuromorphic hardware,'' in \emph{2019 International Joint Conference on
  Neural Networks (IJCNN)}, July 2019, pp. 1--8.

\bibitem{Bartolozzi_Indiveri07a}
C.~Bartolozzi and G.~Indiveri, ``Synaptic dynamics in analog {VLSI},''
  \emph{Neural Computation}, vol.~19, no.~10, pp. 2581--2603, Oct 2007.

\bibitem{Delbruck_etal10}
T.~Delbruck, R.~Berner, P.~Lichtsteiner, and C.~Dualibe, ``32-bit configurable
  bias current generator with sub-off-current capability,'' in
  \emph{International Symposium on Circuits and Systems, ({ISCAS}), 2010},
  IEEE.\hskip 1em plus 0.5em minus 0.4em\relax Paris, France: IEEE, 2010, pp.
  1647--1650.

\bibitem{MIT-BIH}
\BIBentryALTinterwordspacing
G.~B. Moody and R.~G. Mark, ``{MIT-BIH} arrhythmia database,'' 1992. [Online].
  Available: \url{https://physionet.org/physiobank/database/mitdb/}
\BIBentrySTDinterwordspacing

\bibitem{PhysioNet}
A.~L. Goldberger, L.~A.~N. Amaral, L.~Glass, J.~M. Hausdorff, P.~C. Ivanov,
  R.~G. Mark, J.~E. Mietus, G.~B. Moody, C.-K. Peng, and H.~E. Stanley,
  ``{PhysioBank, PhysioToolkit, and PhysioNet}: Components of a new research
  resource for complex physiologic signals,'' \emph{Circulation}, vol. 101,
  no.~23, pp. e215--e220, 2000 (June 13), circulation Electronic Pages:
  http://circ.ahajournals.org/content/101/23/e215.full PMID:1085218; doi:
  10.1161/01.CIR.101.23.e215.

\bibitem{Maass_etal02}
W.~Maass, T.~Natschl{\"a}ger, and H.~Markram, ``Real-time computing without
  stable states: A new framework for neural computation based on
  perturbations,'' \emph{Neural Computation}, vol.~14, no.~11, pp. 2531--2560,
  2002.

\bibitem{Jaeger02}
H.~Jaeger, \emph{Tutorial on training recurrent neural networks, covering
  {BPPT}, {RTRL}, {EKF} and the" echo state network" approach}.\hskip 1em plus
  0.5em minus 0.4em\relax GMD-Forschungszentrum Informationstechnik, 2002.

\bibitem{Langton90}
C.~Langton, ``Computation at the edge of chaos: Phase transitions and emergent
  computation,'' \emph{Physica D: Nonlinear Phenomena}, vol.~42, no.~1, pp.
  12--37, 1990.

\bibitem{aceituno_17}
\BIBentryALTinterwordspacing
P.~V. Aceituno, Y.~Gang, and Y.~Liu, ``Tailoring artificial neural networks for
  optimal learning,'' \emph{CoRR}, vol. abs/1707.02469, 2017. [Online].
  Available: \url{http://arxiv.org/abs/1707.02469}
\BIBentrySTDinterwordspacing

\bibitem{deneve_13}
M.~Boerlin, C.~K. Machens, and S.~Den{\`e}ve, ``Predictive coding of dynamical
  variables in balanced spiking networks,'' \emph{PLoS computational biology},
  vol.~9, no.~11, p. e1003258, 2013.

\bibitem{Alemi_etal2018}
A.~Alemi, C.~K. Machens, S.~Deneve, and J.-J. Slotine, ``Learning nonlinear
  dynamics in efficient, balanced spiking networks using local plasticity
  rules,'' in \emph{Thirty-Second AAAI Conference on Artificial Intelligence},
  2018.

\bibitem{wilten_clopath17}
W.~Nicola and C.~Clopath, ``Supervised learning in spiking neural networks with
  force training,'' \emph{Nature communications}, vol.~8, no.~1, p. 2208, 2017.

\bibitem{Rumelhart_etal86a}
D.~E. Rumelhart, G.~E. Hintont, and R.~J. Williams, ``Learning representations
  by back-propagating errors,'' \emph{Nature}, vol. 323, no. 6088, pp.
  533--536, 1986.

\bibitem{he2019}
X.~He, T.~Liu, F.~Hadaeghi, and H.~Jaeger, ``Reservoir transfer on analog
  neuromorphic hardware,'' in \emph{2019 9th International IEEE/EMBS Conference
  on Neural Engineering (NER)}.\hskip 1em plus 0.5em minus 0.4em\relax IEEE,
  2019, pp. 1234--1238.

\bibitem{Cortes_Vapnik95}
C.~Cortes and V.~Vapnik, ``Support-vector networks,'' \emph{Machine learning},
  vol.~20, no.~3, pp. 273--297, 1995.

\bibitem{rumelhart1985}
D.~E. Rumelhart, G.~E. Hinton, and R.~J. Williams, ``Learning internal
  representations by error propagation,'' California Univ San Diego La Jolla
  Inst for Cognitive Science, Tech. Rep., 1985.

\bibitem{yang_etal11}
{Xiao Yang}, {Qi Cheng}, {Li-fei Lin}, {Wei-wei Huang}, and {Chao-dong Ling},
  ``Design of low power low noise amplifier for portable electrocardiogram
  recording system applications,'' in \emph{2011 IEEE International Conference
  on Anti-Counterfeiting, Security and Identification}, June 2011, pp. 89--92.

\bibitem{ghamati_nejad13}
M.~{Ghamati} and M.~{Maymandi-Nejad}, ``A low-noise low-power {MOSFET} only
  electrocardiogram amplifier,'' in \emph{2013 21st Iranian Conference on
  Electrical Engineering (ICEE)}, May 2013, pp. 1--5.

\bibitem{kumar2016}
R.~Kumar, S.~Sharma, and R.~Goyal, ``A low power low-noise low-pass filter for
  portable {ECG} detection system,'' in \emph{IJCTA}.\hskip 1em plus 0.5em
  minus 0.4em\relax International Science Press, 2016, vol.~9, no.~41, pp.
  95--103.

\bibitem{yehoshuva_etal16}
C.~{Yehoshuva}, R.~{Rakhi}, D.~{Anto}, and S.~{Kaurati}, ``0.5 {V}, ultra low
  power multi standard {Gm-C} filter for biomedical applications,'' in
  \emph{2016 IEEE International Conference on Recent Trends in Electronics,
  Information Communication Technology (RTEICT)}, May 2016, pp. 165--169.

\bibitem{Lichtsteiner_etal08}
P.~Lichtsteiner, C.~Posch, and T.~Delbruck, ``A 128x128 120 d{B} 15 {$\mu$}s
  latency asynchronous temporal contrast vision sensor,'' \emph{{IEEE} Journal
  of Solid-State Circuits}, vol.~43, no.~2, pp. 566--576, Feb 2008.

\bibitem{Qiao_etal15}
N.~Qiao, H.~Mostafa, F.~Corradi, M.~Osswald, F.~Stefanini, D.~Sumislawska, and
  G.~Indiveri, ``A reconfigurable on-line learning spiking neuromorphic
  processor comprising 256 neurons and {128K} synapses,'' \emph{Frontiers in
  Neuroscience}, vol.~9, no. 141, pp. 1--17, 2015.

\bibitem{tibshirani96}
\BIBentryALTinterwordspacing
R.~Tibshirani, ``Regression shrinkage and selection via the lasso,''
  \emph{Journal of the Royal Statistical Society. Series B (Methodological)},
  vol.~58, no.~1, pp. 267--288, 1996. [Online]. Available:
  \url{http://www.jstor.org/stable/2346178}
\BIBentrySTDinterwordspacing

\bibitem{Thakur_etal18}
\BIBentryALTinterwordspacing
C.~S. Thakur, J.~L. Molin, G.~Cauwenberghs, G.~Indiveri, K.~Kumar, N.~Qiao,
  J.~Schemmel, R.~Wang, E.~Chicca, J.~Olson~Hasler, J.-s. Seo, S.~Yu, Y.~Cao,
  A.~van Schaik, and R.~Etienne-Cummings, ``Large-scale neuromorphic spiking
  array processors: A quest to mimic the brain,'' \emph{Frontiers in
  Neuroscience}, vol.~12, p. 891, 2018. [Online]. Available:
  \url{https://www.frontiersin.org/article/10.3389/fnins.2018.00891}
\BIBentrySTDinterwordspacing

\bibitem{zhu2016}
C.~Zhu, S.~Han, H.~Mao, and W.~J. Dally, ``Trained ternary quantization,''
  \emph{arXiv preprint arXiv:1612.01064}, 2016.

\bibitem{hubara2017}
I.~Hubara, M.~Courbariaux, D.~Soudry, R.~El-Yaniv, and Y.~Bengio, ``Quantized
  neural networks: Training neural networks with low precision weights and
  activations,'' \emph{The Journal of Machine Learning Research}, vol.~18,
  no.~1, pp. 6869--6898, 2017.

\bibitem{Helwegen_etal19}
\BIBentryALTinterwordspacing
K.~Helwegen, J.~Widdicombe, L.~Geiger, Z.~Liu, K.~Cheng, and R.~Nusselder,
  ``Latent weights do not exist: Rethinking binarized neural network
  optimization,'' \emph{CoRR}, vol. abs/1906.02107, 2019. [Online]. Available:
  \url{http://arxiv.org/abs/1906.02107}
\BIBentrySTDinterwordspacing

\end{thebibliography}

\end{document}